\documentclass{article}

\usepackage{arxiv}

\usepackage[utf8]{inputenc} 
\usepackage[T1]{fontenc}    
\usepackage{hyperref}       
\usepackage{url}            
\usepackage{booktabs}       
\usepackage{amsfonts}       
\usepackage{nicefrac}       
\usepackage{microtype}      
\usepackage{lipsum}		
\usepackage{graphicx}
\usepackage{natbib}
\usepackage{doi}

\title{Beyond Serendipity: From Exposing the Unknown to Fostering Engagement through Peer Recommendation}

\author{
  \href{https://orcid.org/0009-0002-9226-2341}{\includegraphics[scale=0.06]{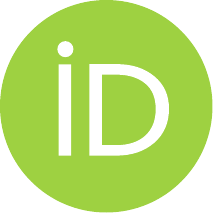}\hspace{1mm}Sosui Moribe} \\
  Graduate School of Design\\
  Kyushu University\\
  Fukuoka, Japan \\
  \texttt{moribe.sosui.695@s.kyushu-u.ac.jp} \\
  \And
  \href{https://orcid.org/0000-0002-1306-0526}{\includegraphics[scale=0.06]{orcid.pdf}\hspace{1mm}Taketoshi Ushiama} \\
  Faculty of Design\\
  Kyushu University\\
  Fukuoka, Japan \\
  \texttt{ushiama@design.kyushu-u.ac.jp} \\
}

\renewcommand{\shorttitle}{Beyond Serendipity: Peer Recommendation}

\hypersetup{
pdftitle={Beyond Serendipity: From Exposing the Unknown to Fostering Engagement through Peer Recommendation},
pdfsubject={q-bio.NC, q-bio.QM},
pdfauthor={Sosui Moribe, Taketoshi Ushiama},
pdfkeywords={First keyword, Second keyword, More},
}

\begin{document}
\maketitle

\begin{abstract}
	Serendipity-oriented recommender systems expose users to unfamiliar items to counter filter bubbles, yet mere exposure does not ensure that users will understand or appreciate the content they encounter. We propose Peer Recommendation, a framework in which a user and an AI agent (Peer) with distinct preferences collaboratively explore unfamiliar content. Unlike conventional conversational recommender systems where the user is a passive recipient, our framework positions the user as both a recommender and a recipient: the user and the Peer mutually recommend songs to each other through chat-based dialogue, collaboratively building a shared playlist. In an exploratory within-subjects experiment (N=14), we compared three conditions: (1) a Close Peer, (2) a Distant Peer, and (3) a baseline agent without an explicit preference profile. The Close Peer significantly increased users' interest expansion and perceived value of the activity compared to the baseline, with medium-to-large effect sizes. The Distant Peer showed no significant difference at the aggregate level; however, qualitative analysis revealed varied responses, with some participants strongly preferring the Distant Peer. These findings suggest that the “otherness” of a recommendation partner is essential for moving beyond mere exposure toward genuine engagement, and that the appropriate degree of preference distance may vary and need to be adapted to individual users.
\end{abstract}


\section{Introduction}
Recommender systems play a central role in modern digital life, supporting everyday decisions about what content to discover and consume across domains such as music, news, and e-commerce. The dominant paradigm optimizes recommendations to match individual user preferences, leveraging collaborative filtering and content-based methods to predict items a user is likely to enjoy. This approach has proven effective at helping users find content that aligns with their established tastes.

However, this over-optimization has been shown to limit the range of content users encounter, leading to filter bubbles and echo chambers. To address this, research on beyond-accuracy objectives has proposed metrics such as diversity and serendipity, aiming to expose users to unfamiliar content \cite{kaminskas2016beyond,kunaver2017diversity}. Yet simply exposing users to unfamiliar items does not ensure that they will understand or appreciate them. What is lacking is a mechanism that supports the process of accepting and appreciating unfamiliar content, not merely exposing users to it.

\begin{figure}
    \centering
    \includegraphics[width=\linewidth]{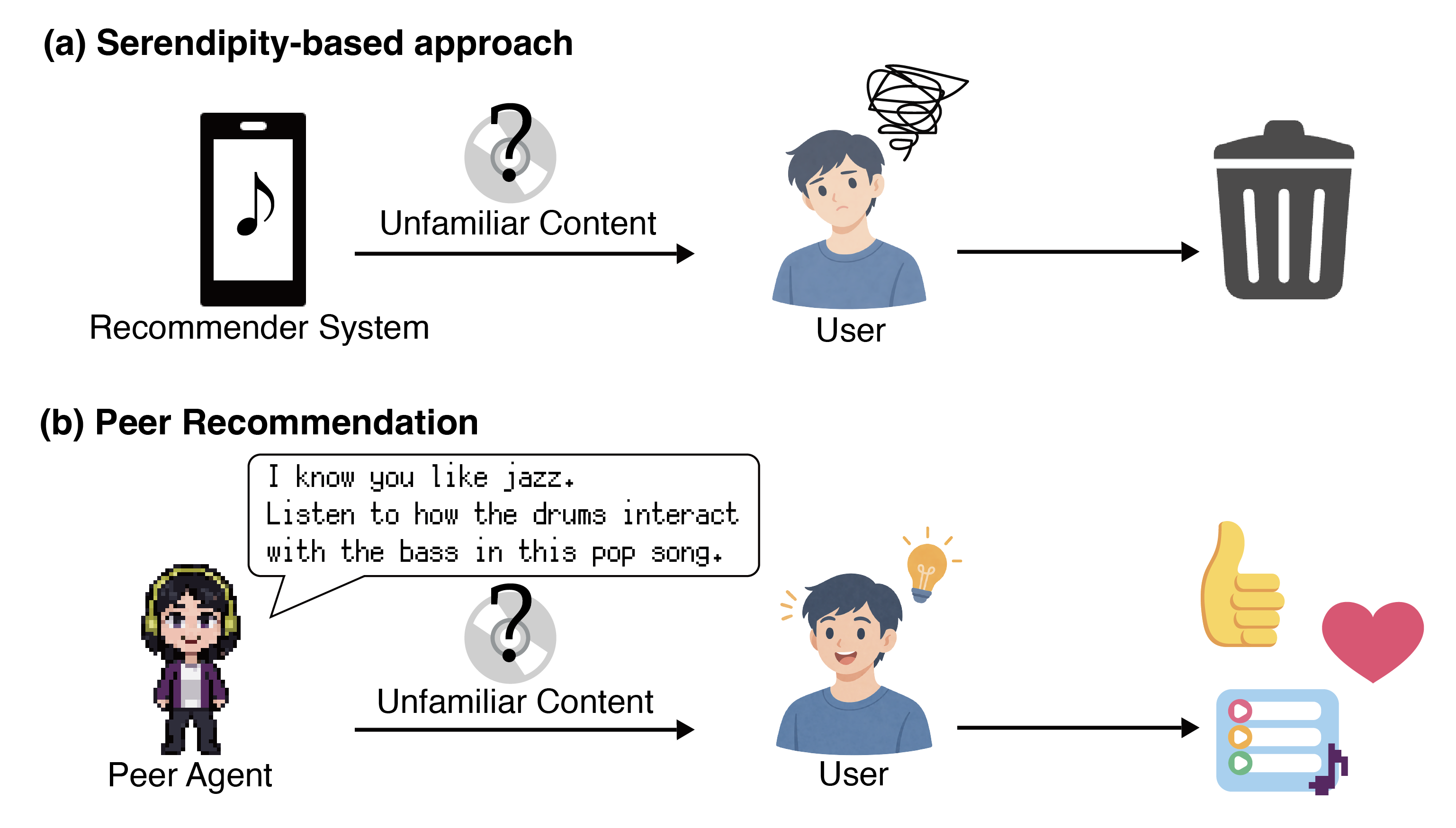}
    \caption{Concept image of Peer Recommendation}
    \label{fig:Concept}
\end{figure}

This study proposes another perspective on recommendation: beyond presenting items that match user preferences, we frame recommendation as a collaborative process that promotes users' active exploration and engagement with unfamiliar content.
This concept draws on Peer Learning \cite{topping2005trends}, a well-established educational framework in which learners with comparable expertise deepen their understanding through mutual explanation and discussion. A conceptual diagram of the concept is shown in Figure~\ref{fig:Concept}. Consider, for example, when friends recommend music or movies to each other: both the recommender and the recipient gain insights by articulating their reasons and engaging with the other's perspective. We term this framework of mutual recommendation "Peer Recommendation" and propose a dialogue-based recommendation environment in which an AI agent functions as a Peer.

We instantiate Peer Recommendation in the domain of music recommendation as a concrete testbed, though the framework is domain-agnostic by design. Our system first constructs a user profile describing the user's musical preferences and values based on their existing playlists and demographic information. From this profile, the system generates a Peer agent with its own distinct preferences, values, and personality traits. Users and Peers then recommend songs to each other through chat-based dialogue and collaboratively build a shared playlist. To examine how preference distance between a user and their Peer affects the user experience, we conducted a within-subjects experiment with 14 participants under three conditions: (1) a Peer with close preferences, (2) a Peer with distant preferences, and (3) a baseline agent with no explicit profile.

Our experiment showed that Peer agents with explicit profiles significantly improved users' perceived interest expansion and perceived value of the activity compared to the baseline agent. 
Notably, the baseline was criticized as "accommodating," with participants reporting that their recommendations were accepted uncritically, indicating that an agent's "otherness"—its capacity to hold and express distinct perspectives—is essential for meaningful collaborative recommendation. 
However, marked individual differences in receptivity to Peer preference distance were observed, suggesting the need for adaptive preference distance control. 
Our contributions are twofold: (1) a framework that reframes recommendation as a collaborative process that fosters active exploration; and (2) we provide empirical evidence that explicit persona design, and the resulting "otherness" of the agent, significantly shapes the quality of collaborative recommendation experiences.

\section{Related Work}
\subsection{Beyond-Accuracy Objectives in Recommendation}
The over-specialization problem in recommender systems, where the range of content users encounter becomes increasingly narrow, has long been recognized. Over the past two decades, a considerable body of research has developed around "beyond-accuracy" objectives such as diversity, serendipity, novelty, and coverage, as surveyed by \cite{kaminskas2016beyond} and \cite{kunaver2017diversity} . These approaches seek to improve recommendation quality by, for example, increasing the dissimilarity among recommended items or promoting less popular content. However, they remain focused on optimizing list-level properties and do not address the process by which users come to understand and appreciate unfamiliar content. Our work takes a fundamentally different approach: rather than improving what is recommended, we redesign how recommendation occurs—through collaborative dialogue with an AI agent that holds distinct perspectives.
\subsection{Conversational Recommender Systems}
Conversational recommender systems, which progressively elicit user preferences through dialogue, have been an active area of research \cite{jannach2021survey, zhang2018conversational}. More recently, LLM-powered approaches have further advanced this paradigm. However, in most of these systems, dialogue serves primarily as a means to clarify user intent and efficiently identify items matching their preferences. The dialogue is not designed as a collaborative process involving mutual understanding between the user and a recommendation partner. In contrast, our work repositions dialogue as the core mechanism for collaborative exploration through mutual recommendation between the user and a Peer agent.

\subsection{LLM-Based Agents with Personas}
Research on building agents with explicit personalities and values using LLMs has gained significant attention. Generative Agents \cite{park2023generative} introduced agents with long-term memory and self-reflection mechanisms that enable coherent, consistent behavior. CharacterLLM \cite{shao2023character} demonstrated an approach for endowing agents with specific personality traits through prompting and training. In the context of recommendation, Shu et al.'s RAH framework \cite{shu2024rah} proposed an LLM-based assistant that learns a user's personality to mediate the recommendation process. A common thread in these works is that agent personas are designed to align with or replicate user characteristics. Our work differs in emphasis: rather than designing the agent's persona to align with the user, we deliberately construct a Peer agent whose preferences and values diverge from the user's, and examine how this "otherness" shapes the collaborative recommendation experience.

\section{Peer Recommendation}
\subsection{Concepts}
Peer Recommendation is a framework in which a user and a Peer agent recommend items to each other on equal footing through dialogue, collaboratively building a shared collection of mutually endorsed items. This framework draws on Peer Learning \cite{topping2005trends}, a well-established approach in educational research where learners deepen their understanding by articulating their own perspectives and contrasting them with those of others. We apply this principle to recommendation and design an environment that promotes users' interest expansion through dialogue-based mutual recommendation, in which both parties articulate their reasons for recommending items and provide feedback on each other's suggestions.
\subsection{System Design}
\begin{figure}
    \centering
    \includegraphics[width=\linewidth]{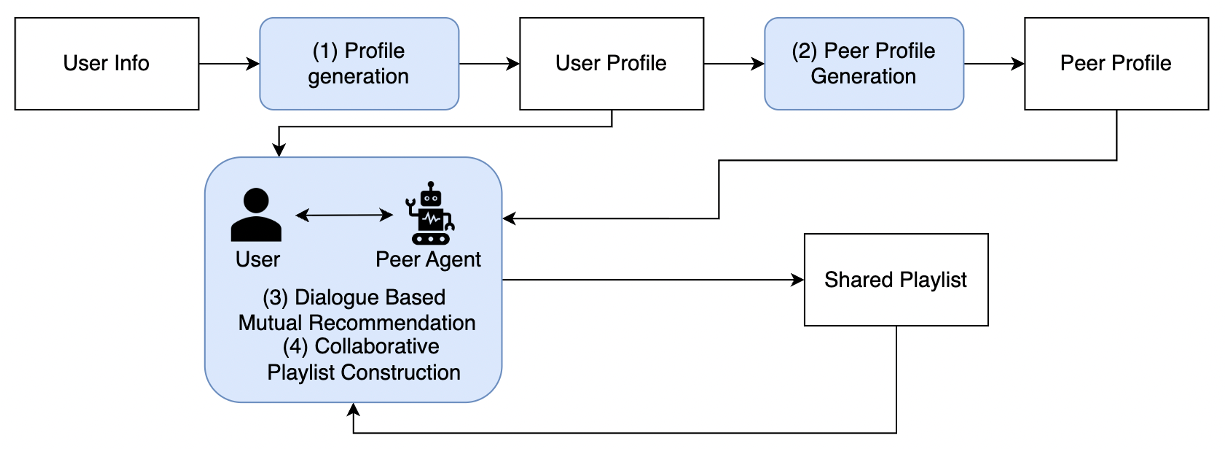}
    \caption{Overview of the Peer Recommendation system pipeline. The system first generates a user profile from basic attributes and existing playlists, then constructs a Peer agent profile (Close or Distant) from the user profile. Users and Peers then engage in mutual recommendation via chat-based dialogue, collaboratively building a shared playlist.}
    \label{fig:system}
\end{figure}

The proposed system consists of four stages (Figure~\ref{fig:system}).

\textbf{(1) User Profile Generation. }Using the user's basic attributes (age, instrument experience, etc.) and existing playlists as input, an LLM generates a profile describing the user's musical values, behavioral tendencies, social listening patterns, and pickiness in natural language.

\textbf{(2) Peer Profile Generation.} From the user profile, the LLM generates a profile for a fictional Peer with distinct preferences, values, and personality traits. To examine the impact of preference distance on the user experience, we designed two types of Peers:
\begin{itemize}
\item \textbf{Type A (Close):} Shares the user's core musical values but holds different perspectives in adjacent domains, intended to encourage deeper exploration of existing preferences.
\item \textbf{Type B (Distant):} Draws primarily from a different preference domain but maintains connection points through shared musical characteristics (e.g., structure, timbre, rhythm), intended to promote boundary expansion.
\end{itemize}

\begin{figure}
    \centering
    \includegraphics[width=\linewidth]{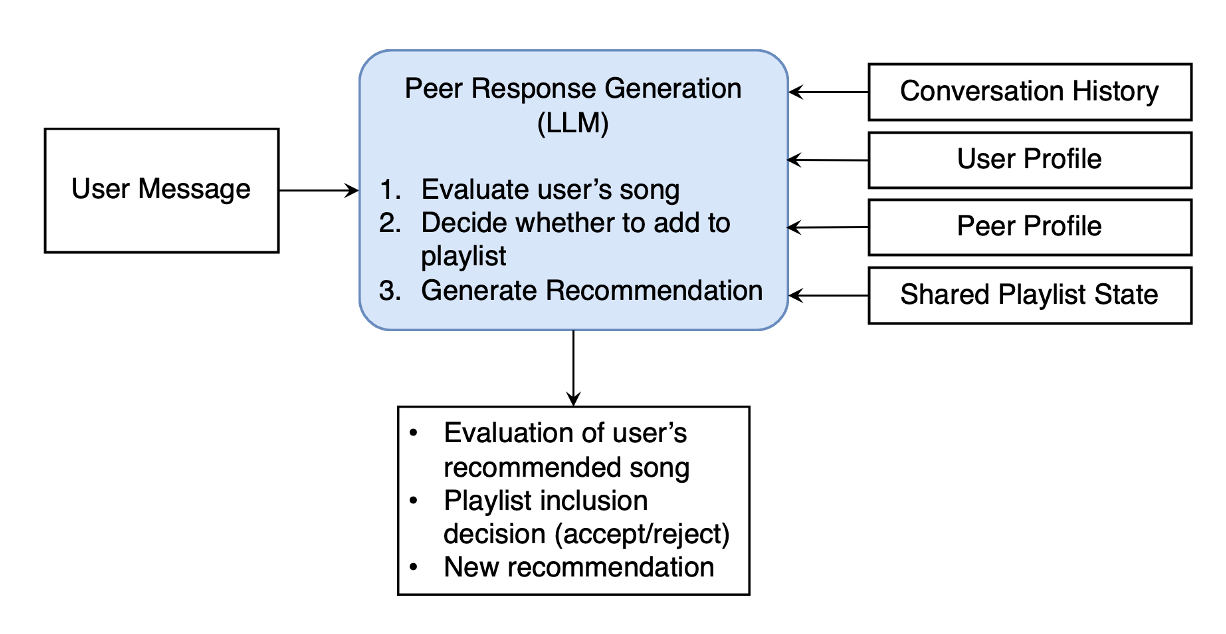}
    \caption{Conversation flow for mutual recommendation. Each user message triggers a three-step pipeline: (1) analysis of user intent and Peer reception, (2) selection of the Peer's next action from candidates, and (3) generation of the Peer's response incorporating the Peer's profile. The Peer profile varies across conditions (Close Peer, Distant Peer, Baseline), shaping how the Peer interprets the user's input and what it recommends in return.
    }
    \label{fig:Conversation Flow}
\end{figure}

\textbf{(3) Mutual Recommendation via Dialogue.} The user and the Peer recommend songs to each other through chat-based dialogue, articulating their reasons and sharing impressions. To generate each Peer response, the system integrates the conversation history, user profile, Peer profile, and the current state of the shared playlist in a single LLM call, producing an evaluation of the user's recommended song, a decision on playlist inclusion, and a new recommendation. The flowchart for this section is shown in Figure~\ref{fig:Conversation Flow}.

\textbf{(4) Collaborative Playlist Construction.} The recipient of each recommendation decides whether to add the song to the playlist; only songs positively evaluated by both parties are included. This process yields a playlist collaboratively constructed through mutual agreement.

We note that both user and Peer profiles are fixed at the start of each session and are not updated during the dialogue. This design choice eliminates confounding effects of profile updates, allowing us to isolate the effect of the Peer's preference distance on the user experience.

\section{Experiment}
\subsection{Experimental Design}
We conducted a within-subjects user experiment with 14 participants. Each participant interacted with a Peer agent through chat-based dialogue, mutually recommending songs and collaboratively building a shared playlist. Each participant experienced all three conditions:
\begin{itemize}
\item\textbf{Close Peer:} A Peer with preferences and values close to the user's.
\item\textbf{Distant Peer:} A Peer with preferences significantly different from the user's.
\item\textbf{Baseline:} An LLM agent with no explicit preference profile.
\end{itemize}
The Baseline condition lacks an explicit preference profile, causing it to function as a general-purpose assistant that tends to accommodate user preferences rather than express independent perspectives. Since the underlying response generation framework is otherwise identical across all conditions, this design isolates the contribution of the explicit preference profile to the collaborative recommendation experience.
We assessed the impact of Peer preference distance on user experience using multiple measures: a single-item measure of perceived interest expansion and subscales of the Intrinsic Motivation Inventory (IMI) including value and tension/pressure. 
We also collected open-ended feedback for each condition. 
Based on our hypothesis that Peer agents with explicit profiles would increase users' perceived interest expansion and perceived value compared to the baseline, we employed one-tailed Wilcoxon signed-rank tests for these directional comparisons (two-tailed for Tension/Pressure). Effect sizes were computed as $r = Z/\sqrt{N}$, where N denotes the total sample size.

\subsection{Quantitative Results}
Table~\ref{tab:results} summarizes the main results.
 
\begin{table}[h]
\centering
\caption{Main quantitative results. 
CP = Close Peer, DP = Distant Peer, BL = Baseline. 
One-tailed Wilcoxon signed-rank tests were used for 
directional hypotheses; two-tailed tests were used for 
Tension/Pressure. Bold $p$-values indicate statistical 
significance ($p < .05$). $^\dagger$ indicates a trend 
toward significance ($p < .10$).}
\label{tab:results}
\small
\begin{tabular}{lccccc}
\toprule
Measure & CP & DP & BL & CP vs BL & DP vs BL \\
 & $M$ & $M$ & $M$ & ($p$, $r$) & ($p$, $r$) \\
\midrule
Interest expansion & 5.21 & 4.64 & 4.43 
  & \textbf{.044}, .46 & .453, .03 \\
IMI: Value & 5.62 & 4.75 & 4.55 
  & \textbf{.006}, .67 & .431, .05 \\
IMI: Tension/Pressure & 2.04 & 2.21 & 2.29 
  & 1.00, .00 & .865, .05 \\
\bottomrule
\end{tabular}
\end{table}

\textbf{Interest expansion.}
On the single-item measure of perceived interest expansion, 
Close Peer ($M=5.21$) was significantly higher than 
Baseline ($M=4.43$), with a medium-to-large effect size 
($p=.044$, $r=.46$). The distribution of scores across 
conditions is shown in Figure~\ref{fig:boxplot}.

\begin{figure}[t]
  \centering
  \includegraphics[width=\columnwidth]{combined_boxplot.png}
  \caption{Box plots of perceived interest expansion (left) 
  and IMI: Value (right) across conditions. Horizontal lines 
  indicate medians; diamonds indicate means. Close Peer shows 
  both higher means and lower variance than Baseline on both 
  measures.}
  \label{fig:boxplot}
\end{figure}

\textbf{IMI subscales.}
On the Value subscale, Close Peer ($M = 5.62$) was significantly higher than Baseline ($M = 4.55$), with a large effect size ($p = .006$, $r = .67$).
The distribution of Value scores is also shown in Figure~\ref{fig:boxplot}.
Tension/Pressure remained low across all conditions (Close Peer ($M=2.04$), Distant Peer ($M=2.21$), Baseline ($M=2.29$)), confirming that none of the conditions imposed undue psychological burden.
 
\textbf{Variance and correlation.}
The variance in ratings for Close Peer and Distant Peer was 
approximately half or less than that of Baseline across 
multiple measures (one-tailed Pitman-Morgan test: interest 
expansion, Close Peer vs.\ Baseline: $p = .028$, Distant Peer vs.\ Baseline: $p = .039$; 
IMI Value, Close Peer vs.\ Baseline: $p = .004$), indicating that the proposed conditions provided a more consistent experience.

No significant correlation was found between subjective preference 
alignment and perceived interest expansion ($r_s = .19$, $p = .51$), 
suggesting that the perceived similarity to the Peer is not a primary 
driver of interest expansion. 
The lack of a significant effect for Distant Peer at the aggregate level may reflect varied responses among participants; we explore this in the qualitative analysis below.

\subsection{Qualitative Analysis}
Analysis of open-ended comments revealed marked individual 
differences in receptivity to Peer preference distance. As a 
post-hoc analysis, we identified two participant types based 
on their evaluation patterns.

\textbf{Exploration-oriented participants} viewed recommendations 
that deviated from their preferences positively and rated the 
Distant Peer highly. In contrast, they criticized 
Baseline as overly accommodating, reporting that they 
"felt no pushback" and that "no discussion was sparked." 
These comments were directed at the baseline agent's tendency 
to accept the user's recommendations almost uncritically, 
suggesting that an agent's "otherness"---its capacity to 
express distinct perspectives and disagreement---is important 
for meaningful collaborative recommendation.

\textbf{Affinity-oriented participants} valued recommendations 
close to their preferences as providing a "sense of being 
understood" and preferred the Close Peer and 
the Baseline. Recommendations from the Distant 
Peer tended to be perceived as cognitive burden.

Notably, some participants rated Distant Peer higher than Close Peer. 
For instance, one participant scored the conditions Close Peer:5, Distant Peer:7, Baseline:1, commenting that Distant Peer recommended songs from culturally distant genres that nonetheless resonated deeply, while describing Baseline as failing to spark any new discovery. 
Another participant (Close Peer:2, Distant Peer:6, Baseline:2) noted that the Distant Peer "pushed its own interests rather than accommodating mine," contrasting this with Baseline's lack of novelty. 
These cases suggest that for some users, preference distance serves as a catalyst for discovery, further supporting the need for adaptive 
control of Peer preference distance.

\subsection{Discussion}
Our results suggest that the effect of Peer preference distance varies depending on users' exploration orientation. 
This implies that a one-size-fits-all approach to preference distance is insufficient. 
In a practical deployment, users could be given control over how far the Peer's preferences diverge from their own, allowing them to calibrate the balance between familiarity and challenge according to their individual readiness for exploration.

\section{Conclusion}
This study proposed Peer Recommendation, a framework that reframes recommendation as a collaborative process for fostering users' active exploration and perceived value of recommendation activities, and evaluated a dialogue-based environment for music recommendation.
Our experiment showed that a Peer agent with an explicit preference profile significantly improved users' perceived interest expansion and perceived value of the activity compared to a baseline agent without a profile.
Notably, the baseline agent was criticized as "accommodating," highlighting the importance of the agent functioning as an "other" with distinct perspectives and values in conversational recommendation.
However, marked individual differences in receptivity to Peer preference distance were observed, suggesting the need for adaptive control of preference distance tailored to individual users.
While we used music recommendation as a testbed, the framework generalizes to any domain where collaborative list-building occurs, such as curating reading lists, news bookmarks, or learning resources.
Prompts will be made available upon request.

\section*{Acknowledgements}
This work was supported by JST BOOST, Japan Grant Number JPMJBS2406.
We thank all participants for their time and valuable feedback.

\bibliographystyle{unsrt}
\bibliography{references}  






\end{document}